# Automated Sustainability Compliance Checking Using Process Mining and Formal Logic

Clemens Schreiber
AIFB, Karlsruhe Institute of Technology, Karlsruhe, Baden-Württemberg, Germany
clemens.schreiber@kit.edu

**Abstract**
Business processes need to have certain constraints such that they can lead to sustainable outcomes. These constraints can be manifold and their adherence has to be monitored. In the past compliance checking has been applied in several business domains without considering certain sustainability aspects, such as multi-dimensionality and impact level. With my research I want to contribute to the application of compliance checking techniques for the purpose of sustainability compliance. In order to achieve this, I want to analyse and develop data-driven approaches, which allow to automate the task of compliance checking. The way in which this can be achieved, is be combining methods from process mining with formal languages that can express sustainability rules in a machine-readable manner. The main goal is to develop a compliance engine that can be adapted by ERP systems in order to evaluate sustainability conformance in business processes.

**Motivation and Research Objective**
In the recent past sustainability considerations have also become an important part of business conduct. The European Union introduced in 2014 new regulations for environmental, social and governance disclosures [1], and in 2015 policies for a Circular Economy [2]. At the same time there also exist further initiatives such as self-regulations, voluntary environmental agreements, and public-private partnerships [1]. It is in the public interest, that companies follow certain standards and regulations, such that their business activities lead eventually to beneficial outcomes for company stakeholders. Two examples for such regulations are the Sarbanes-Oxly Act of 2002, regulating the disclosure of corporate finance reports to the public, and Basel III, regulating the risk management in the banking sector.

But in order to ensure that such regulations and policies are also enforced in business practice, business activities need to be regularly monitored and evaluated. Compliance checking can help to monitor business processes and to prevent misconduct. In [2] five types of compliance-related activities are identified: compliance elicitation, compliance formalization, compliance implementation, compliance checking, and compliance improvement. The first two activities consider how regulations or policies can be reflected in process constraints and how they can be expressed. This information is used in the phase, to configure information systems such that they adhere to the identified process constraints. The compliance checking tries in a next step to verify whether the process is compliant with the given constraints or not. In this way further information is generated, which can be used for process improvement.

The compliance checking can be conducted either as (1) forward compliance checking (before the process execution) or (2) backward compliance checking (after the process execution) [2]. For the latter I will propose a data-driven approach, which can help to automate the compliance checking process and increase its efficiency. The main purpose of my research will be to identify and to extend existing compliance checking methods for sustainability constraints in business processes. In comparison to the existing methods, sustainability compliance faces some particular challenges, which are for example stated in the Karlskrona Manifesto for Sustainability [3]:

- sustainability has multiple dimensions
- sustainability requires long-term thinking
- the impact on sustainability has to be considered on direct, indirect and systemic level
- sustainability applies to a system as well as to its wider context.

These challenges can make it difficult to define clear sustainability constraints for business processes and make it costly to check for compliance. The overall goal of my research is to facilitate the sustainability compliance checking for business processes by automating the monitoring process. For the automation of the compliance checking one has to provide an information system with formal (machine-readable) sustainability rules, the business process execution data, and context data to measure the impact on the environment. Two main challenges for this approach are to find ways of defining formal sustainability rules and to create accurate context models that can sufficiently represent sustainability aspects.

Once the constraints are defined, different process mining techniques can be used to execute the compliance checking, based on process execution data. With process mining it is possible to derive process models based on the event log and to analyse the overall process behaviour. Given that sufficient data is available for the analysis, this approach can potentially facilitate sustainability compliance checking and make it more efficient.

This leads to the following two main research questions. How can sustainability constraints for business processes be expressed by formal language? How can the formal representation of sustainability constraints for business processes be combined with process mining techniques, in order to automate sustainability compliance checking?

---

[1] Disclosure of non-financial and diversity information. Available online: https://eur-lex.europa.eu/legal-content/EN/TXT/?uri=CELEX%3A32014L0095 (accessed on 13.03. 2020).

[2] Circular Economy Action Plan. Available online: https://eur-lex.europa.eu/legal-content/EN/TXT/?qid=1551871245356&uri=CELEX:52019SC0090 (accessed on 13.03.2020).

## Background

**Automation of Sustainability compliance checking**
Ramezani et al. [2] distinguish between two different approaches to formalize compliance constraints in processes: (1) logical languages (e.g. LTL) and (2) process patterns (e.g. petri-net). (1) Logical languages are able to define structural constraints of a processes, such as the order of activities or the dependency of activities within a process. At the same time, they can control for context parameters, such as resource restrictions. In the case of human resources, these restrictions can for example specify the dependencies between the people involved in a process [4]. To which extend sustainability constraints can be formalized depends on the expressiveness of the logical language [5]. (2) With process patterns it is also possible to express other compliance dimensions besides the control flow, such as data flow (e.g. "Productions with $CO_2$ emissions more than 1 ton, require two validity checks."), and organisational rules (e.g. "Each validity check must be conducted by two different departments.") [2]. However, there are many more properties of processes that can be defined by compliance rule, e.g. time-related constraints.

In addition to the rule-based approaches to sustainability compliance checking, machine learning techniques, such as association rule learning (ARL), can be applied to detect deviations in process behaviour. This was for example successfully applied in the case of fraud detection [6]. The advantage of this method is, that it can identify misconduct, even when there is no direct violation of compliance rules. The disadvantage is that this approach can only be used for backward compliance checking and cannot prevent present misconduct. But if process data already exist ARL can be applied in combination with Process Mining in order to automatically detect deviations in process behaviour. Context-aware Process Mining techniques enable a data driven approach to compliance checking and facilitate the monitoring task. Process mining can be combined with logical language [2], process patterns [2], and machine learning [6].

A main challenge for the automation of sustainability compliance checking is the dependency on process data. This is in particular an important issue in highly opaque fields, where practices, causality, and performance are difficult to assess [7]. In socioenvironmental governance it is for example often not easy to understand the relation between corporate activities and socioenvironmental outcomes. Hence, in order to trace down this relation a vast amount of data would be needed to successfully apply data-driven conformance checking. One possibility to deal with this issue is to use estimation of missing data and to carefully apply assumptions about the relation between a business process and its impact on sustainability.

## Related work

The integration of sustainability aspects in business processes can be a challenging task. On the one hand there are multiple sustainability dimensions to be considered, which are interdependent, and on the other hand it can be difficult to evaluate the impact of process activities on society and environment [7, 8]. Hence, the integration and evaluation of sustainability aspects in business processes can require a lot of effort. One approach to facilitate the integration of sustainability aspects in business processes is the development of sustainability patterns that can be generally applied to multiple business domains [9, 10]. These patterns can be used to improve the process performance in terms of sustainability, by suggesting changes on a structural or on an activity level. Redistributing resources within a business process can for example increase the energy efficiency of a particular activity, while changing the order of activities within a workflow can lead to sustainability improvement on a structural level. Using such general patterns can facilitate the formulation of sustainability rules, that can be applied for automated conformance checking. However, the existing description of sustainability patterns only focuses on the design phase of business processes, not on the execution phase.

The automated execution of sustainability compliance checking has been of minor interest in ICT research so far. This might be the case, because the existing research has considered compliance checking as a general issue, mainly related to finance and auditing and not to sustainability [2, 6, 11, 12].

## Research approach

**Research Goals**
Regulations, which seek to ensure sustainability in businesses processes face some particular issues such as multi-dimensionality, long term perspective, and multi-level impact considerations. How these particular issues can be expressed formally in a feasible manner has yet to be discovered. This will be one main objective of my research. The other objective will be to develop a compliance engine for sustainability compliance checking of business processes, based on formal language. The existing architecture will in particular be enhanced by a process mining module. This module will ensure that structural sustainability constraints of the business process will be fulfilled.

**Research Agenda**
A *first* step of my research is to identify existing sustainability regulations, which can be formalized and used for automated sustainability compliance checking. In order to identify and develop a sufficient representation for sustainability constraints it will be necessary to identify existing regulations for business processes. Depending on the regulations it will be possible to define requirements for the expressiveness of a formal representation of sustainability constraints. The requirement elicitation should also be conducted in cooperation with domain experts in order to ensure the usability of this approach.

After the identification of a suitable formal representation, the *second* research step is to use logical expressions, process patterns and machine learning in order to monitor the adherence of regulations in existing ERP systems for the purpose of automated sustainability compliance checking. Which particular approach will be the most suitable method for sustainability compliance checking will also be evaluated in the first research stage. However, in the second stage there will be a more in-depth analysis, especially in terms of technical feasibility.

The *third* research step will be to develop an adaptable compliance engine, similar to the one described in [13]. The engine has access to the runtime process execution data, which includes process data and (external) context data. It also has accesses to the sustainability regulations defined by the business enterprise. The business process, as well as the sustainability constraints can be constantly adjusted. Whenever the business process behaviour is not compliant with the given constraints, the system will recognize this and inform the management at the business enterprise (see Figure 1).

The *fourth* research step will be, to evaluate this approach in terms of efficiency, flexibility and usability. Although this evaluation will also be conducted during the other research stages, the last stage will in particular look at the applicability of the developed compliance engine. The evaluation at this stage will also consider possible cost reductions, as well as the reliability of the approach.

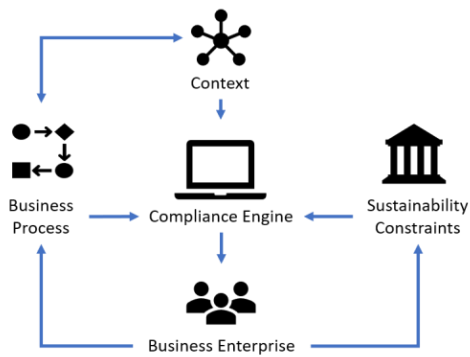

**Figure 1.** The compliance engine considers business process data and context data in order to inform the business enterprise about sustainability compliance.

### Current Research Status

My research on sustainability compliance checking is still in its early beginning. Some technical tests have been conducted in order to see whether the approach is feasible or not. Conceptually the compliance engine in [13] and existing implementations known as Green IS [14] indicate that the combination of different compliance checking methods could be implemented with a feasible effort. Further tests will show how efficient and reliable this approach is.